# Electronic properties across metal-insulator transition in $\beta$-pyrochlore-type CsW$_2$O$_6$ epitaxial films

Takuto Soma[1], Kohei Yoshimatsu*,[1], Koji Horiba[2,3], Hiroshi Kumigashira[2,3], and Akira Ohtomo[1,3]

[1]Department of Chemical Science and Engineering, Tokyo Institute of Technology, 2-12-1 Ookayama, Meguro, Tokyo 152-8552, Japan

[2]Photon Factory, Institute of Materials Structure Science, High Energy Accelerator Research Organization (KEK), 1-1 Oho, Tsukuba 305-0801, Japan

[3]Materials Research Center for Element Strategy (MCES), Tokyo Institute of Technology, Yokohama 226-8503, Japan

**ABSTRACT:** In CsW$_2$O$_6$, which undergoes a metal-insulator transition (MIT) at 213 K, the emergence of exotic properties associated with rattling motion of Cs is expected owing to its characteristic $\beta$-pyrochlore-type structure. However, a hurdle for crystal growth hampers elucidation of detailed properties and mechanisms of the MIT. Here we report on the epitaxial growth of $\beta$-pyrochlore-type CsW$_2$O$_6$ films and their electronic properties across the MIT. Using pulsed-laser deposition technique, we grew single-crystalline CsW$_2$O$_6$ films exhibiting remarkably lower resistivity compared with a polycrystalline bulk and sharp MIT around 200 K. Negative magnetoresistance and positive Hall coefficient were found, which became pronounced below 200 K. The valence-band and core-levels photoemission spectra indicated the drastic changes across the MIT. In the valence band photoemission spectrum, the finite density of states was observed at the Fermi level in the metallic phase. In contrast, an energy gap appeared in the insulating phase. The split of W $4f$ core-level spectrum suggested the charge disproportionation of W$^{5+}$ and W$^{6+}$ in the insulating phase. The change of spectral shape in the Cs $4d$ core levels reflected the rattling motion of Cs$^+$ cations. These results strongly suggest that CsW$_2$O$_6$ is a novel material, in which MIT is driven by the charge disproportionation associated with the rattling motion.

## ■ INTRODUCTION

$\beta$-pyrochlore-type oxides ($AM_2$O$_6$) have recently attracted much attention as they show anomalous phonon induced properties [1-4]. The $\beta$-pyrochlore is also known as defect-type pyrochlore [5]. The crystal structures of normal and $\beta$-pyrochlore-type oxides are schematically shown in Fig. 1. In contrast to the normal pyrochlore ($A'_2M_2$O$_7$) [Fig. 1(a)], the $A$ cations occupy O-absent space inside vast cages surrounded by the octahedral $M$O$_6$ framework, leaving the original $A'$-site entirely empty [Fig. 1(b)]. As a result, anomalous local atomic vibration of the $A$ cations called rattling is realized [1–4]. The rattling phonon is responsible for strong electron-phonon interaction and low thermal conductivity, which trigger exotic properties such as superconductivity and high thermoelectric effects [4, 6, 7].

Among $\beta$-pyrochlore-type oxides, $A$Os$_2$O$_6$ ($A$ = K, Rb, Cs) have been actively investigated. Preparation of high-quality single crystals is possible and various properties have been elucidated [2–4]. For example, $A$Os$_2$O$_6$ show superconductivity associated with rattling motion of the $A$ cations and their superconducting transition temperature varies in accordance with the $A$ atoms.

Except for the osmates, CsW$_2$O$_6$ is the only example of $\beta$-pyrochlore-type oxides having $d$ electrons, which is first synthesized by Cava *et al*. in 1993 [8]. In addition, it is the only example of pyrochlore-type tungstate having $d$ electrons regardless of normal or $\beta$-pyrochlore. However, CsW$_2$O$_6$ is thermally unstable so that with using conventional synthesis methods even a dense polycrystalline sample cannot be prepared, much less a single crystal [8, 9]. Thus, a little are known about the physical properties. In 2013, Hirai *et al*. reported that the MIT occurred at 213 K accompanying with a structural transition (cubic to orthorhombic), and a small magnetization was suddenly quenched in the insulating phase [9]. As they pointed out, the sample quality affected the physical properties. Furthermore, the magnetotransport and electron spectroscopy experiments have not been carried out, yet. Thus, it has been enthusiastically desired to reveal intrinsic properties of CsW$_2$O$_6$ in viewpoints of understanding not only a novel class $\beta$-pyrochlore-type oxide, but also an analogue to 5$d$ pyrochlore $A_2$Ir$_2$O$_7$ ($A$ is lanthanides), which exhibits topological states owing to strong spin-orbit coupling [10–13].

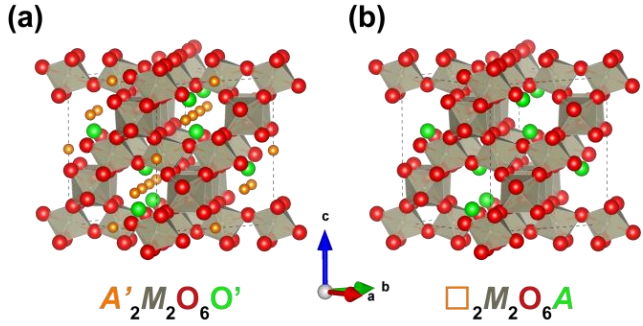

**Figure 1**. Schematic illustrations of the crystal structures of (a) normal pyrochlore-type $A'_2M_2O_7$ and (b) $\beta$-pyrochlore-type $AM_2O_6$. Note that the locations of O' in (a) and $A$ atoms in (b) are identical to each other.

In this study, we have fabricated $\beta$-pyrochlore-type $CsW_2O_6$ epitaxial films to investigate their electronic properties in detail. Using the pulsed-laser deposition (PLD) technique, high-quality single-crystalline $CsW_2O_6$ films were successfully grown on lattice-matched Y-stabilized $ZrO_2$ (YSZ) substrates. Temperature dependence of resistivity indicated the clear MIT around 200 K. From magnetotransport measurements, negative magnetoresistance was observed in lower and higher temperatures near MIT. Synchrotron-radiation photoemission measurements were performed at temperatures over the MIT. The valence band spectra reflected a gap near the Fermi level ($E_F$) in the insulating phase. The W $4f$ core levels evolved their spectral shape and two components were discriminated in the insulating phase, being consistent with charge disproportionation of $W^{5+}$ and $W^{6+}$ states. The Cs $4d$ core-level spectra drastically changed across the MIT, reflecting the rattling motion of $Cs^+$ cations. These results suggest that the MIT of $CsW_2O_6$ originates from the charge disproportionation of W cations and the rattling motion of $Cs^+$ cations.

■ **EXPERIMENTAL SECTION**

$CsW_2O_6$ films were grown on YSZ (111) substrates by using PLD technique with a KrF excimer laser (0.6 J cm$^{-2}$, 6 Hz). A $CsW_2O_{6+\delta}$ ceramic tablet as a laser-ablation target was prepared by conventional solid-state reaction steps, starting from mixture of $Cs_2CO_3$ and $WO_3$ powders with a stoichiometric molar ratio (Cs : W = 1 : 2). Generally, for compounds containing alkali metals, the adjustment of film stoichiometry is often difficult in PLD growth due to high vapor pressure of alkali metals [14]. For the present Cs-W-O system, however, stoichiometric transfer of cation compositions from targets to films has been revealed in our previous study [15]. The substrate temperature and oxygen pressure ($P_{O_2}$) were set to 650 °C and 7.5 mTorr, respectively, while pure oxygen (6 N purity) was fed into a vacuum chamber during the growth. The film thickness ranged from 30 to 200 nm, as verified by a stylus-type profiler. Most of the properties were measured for ~150-nm-thick films.

The structural properties were investigated by a laboratory X-ray diffraction (XRD) apparatus with Cu K$\alpha_1$ radiation. The temperature dependence of resistivity and magnetoresistance were measured by a standard four-probe method using a physical property measurement system (Quantum Design, PPMS). The magnetic field was applied perpendicular to film surface. Ti/Au metal electrodes were vacuum-evaporated on the films for electrical ohmic contacts.

Photoemission spectroscopy (PES) was performed for samples transferred *ex-situ* at the undulator beamline of BL-2A in the Photon Factory, KEK. The PES spectra were recorded using an electron energy analyzer (SES-2002, VG Scienta) with an energy resolution of less than 150 meV at phonon energy of 630 eV at RT. $E_F$ was referred to that of Au electrically in contact with the sample surface.

■ **RESULTS AND DISCUSSION**

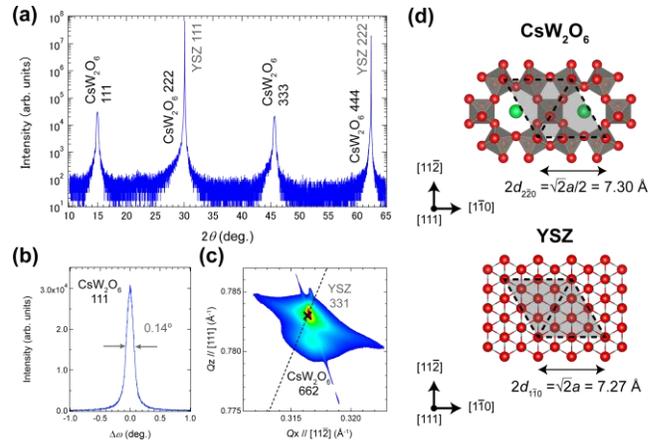

**Figure 2.** (a) Typical out-of-plane XRD pattern for $CsW_2O_6$ films on YSZ (111) substrates grown under the optimized condition. (b) $\omega$-scan for the $CsW_2O_6$ 111 reflection. (c) reciprocal space map around the YSZ 331 reflection. The reciprocal point of bulk $CsW_2O_6$ 662 reflection is indicated by a crossing bars. The broken line intersects the axis origin. (d) Schematics of crystal structures of $CsW_2O_6$ and YSZ projected along [111] directions. The red and green spheres indicate oxygen and Cs atoms, respectively. The hatched areas surrounded by broken lines represent the surface unit cells for epitaxial growth.

We first describe the structural properties of a single-crystalline $CsW_2O_6$ epitaxial film. Figure 2(a) shows the out-of-plane XRD pattern of the $CsW_2O_6$ film. Only $\beta$-pyrochlore-type $CsW_2O_6$ odd *hhh* reflections were detected along YSZ 111 and 222 reflections, indicating (111)-oriented $CsW_2O_6$ films. While $CsW_2O_6$ even *hhh* reflections were hardly discriminated from the substrate reflections due to their close lattice constants to each other [16]. Clear Laue fringes were observed at each reflection peak, suggesting high-quality films. As shown in Fig. 2(b), the full width at half maximum of the $\omega$-scan of the $CsW_2O_6$ 111 reflection was 0.14°, indicating high in-plane lattice coherency. To investigate further details about the in-plane lattice structures, we also took reciprocal space map around the YSZ 331 reflection as shown in Fig. 2(c). The in-plane epitaxial relationship was verified as $CsW_2O_6$ [1$\bar{1}$0] || YSZ [1$\bar{1}$0]. Note that this epitaxial relationship is identical to that of normal pyrochlore-type iridate films grown on YSZ (111) substrates [17, 18]. In addition, $Q_x$ values between the film and substrate were identical, indicating that $CsW_2O_6$ films were coherently grown on the YSZ lattice.

Figure 2 (d) illustrates epitaxial relationship between $CsW_2O_6$ and YSZ projected along the [111] direction. The double $d_{2\bar{2}0}$ spacing of $CsW_2O_6$ (7.30 Å) is close to double $d_{1\bar{1}0}$ spacing of YSZ (7.27 Å), and their mismatch is only 0.4% [9]. Therefore, high-quality single-crystal $CsW_2O_6$ films could be grown on the lattice matched substrates. We would like to emphasize that this is the first demonstration of epitaxial growth of β-pyrochlore-type compounds. We also note that $CsW_2O_6$ bulk obtained previously are low-dense polycrystals, but not single crystals [8, 9]. The non-equilibrium thin-film synthesis and excellent lattice matching to the YSZ substrate realizes epitaxial stabilization of the meta-stable β-pyrochlore phase.

Next, we describe transport properties of the $CsW_2O_6$ film. Figure 3(a) shows the temperature dependence of resistivity $\rho$ for the $CsW_2O_6$ film. For comparison, we also plotted the $\rho$-$T$ curve of low-dense polycrystalline bulk $CsW_2O_6$ [9]. The room-temperature $\rho$ of our film was ~$10^{-2}$ Ω cm, which was lower by two orders of magnitude than that of bulk. The lower $\rho$ is certainly attributed to high crystallinity. Below 200 K, resistivity indicated sharp increase with narrow hysteresis, which was also found in bulk and thought to be the sign of the MIT [9]. The existence of the hysteresis implies the occurrence of the charge disproportionation as is reported in various mixed-valent compounds, which will be discussed in connection with core-level measurements later [19]. Slight difference in transition temperature (200 K in our film and 213 K in bulk) can be attributed to epitaxial strain and/or slight off-stoichiometry as reported for various compounds exhibiting the MIT [20–22].

In previous study for bulk, high-temperature (HT) phase ($T$ > 213 K) is assigned to be metallic despite d$\rho$/d$T$ < 0, and the opposite d$\rho$/d$T$ sign is ascribed to low crystalline quality that hinders the intrinsic properties [9]. On the other hand, our film also showed an insulating behavior despite high crystalline quality. This fact suggests that bad metallicity in the HT phase is intrinsic to $CsW_2O_6$. It should be mentioned that although crystal quality was independent of film thickness, the MIT disappeared when film thickness was reduced down to several tens of nanometers.

In low-temperature (LT) phase, conduction mechanism was different from the previous reports. In Ref. 13, $\rho$-$T$ in a LT range is described with thermally activated conduction: $\rho(T) = \rho_o \exp(E_a/k_BT)$, where $\rho_o$ is the preexponential term, $E_a$ is activation energy, and $k_B$ is Boltzmann's constant.

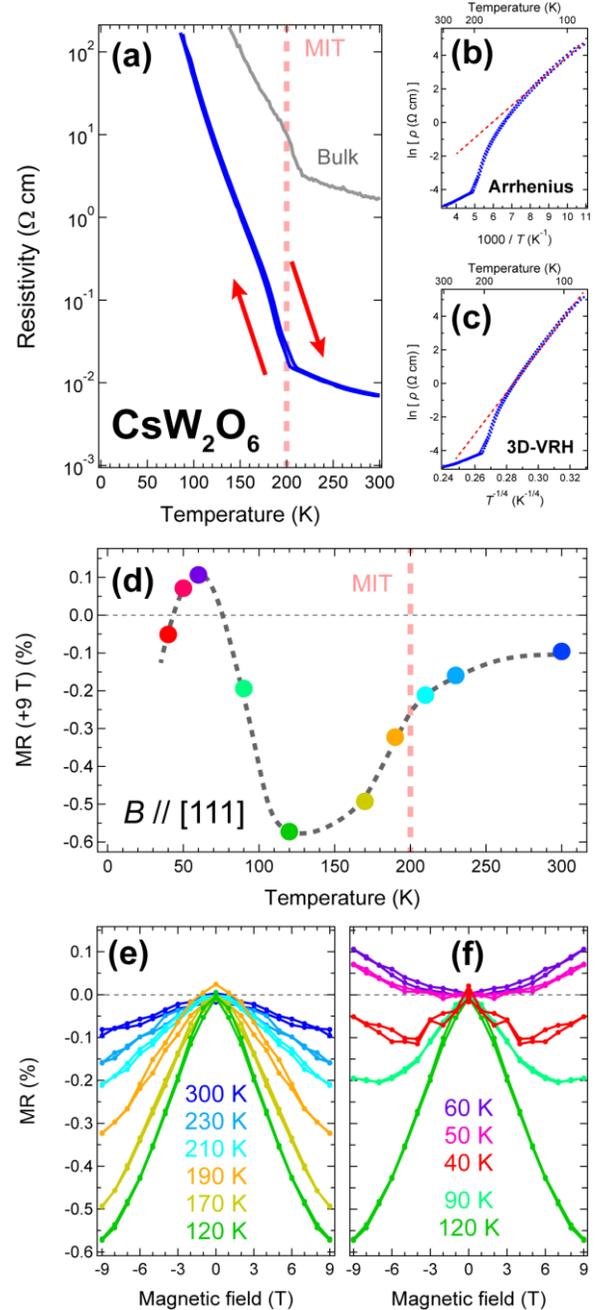

**Figure 3.** (a) Temperature dependence of the resistivity for the $CsW_2O_6$ film. The bulk data are also plotted as a reference [9]. Red arrows and broken lines represent the direction of the temperature sweep and the transition temperature for the film, respectively. (b) ln $\rho$ vs. 1000/$T$. The broken line is a linear fit to thermal activation model. (c) ln $\rho$ vs. $T^{-1/4}$. The broken line is a linear fit to VRH model. (d) Temperature dependence of the magnetoresistance for the CsW$_2$O$_6$ film under 9 T. The gray broken line indicates a guide to the eye. (e) and (f) Magnetic field dependence of the magnetoresistance taken at different temperature ranges.

In contrast, the LT transport properties of our film were better described with 3-dementional variable-range hopping (VRH) conduction: $\rho(T) = \rho_o \exp(T_o/T)^{1/4}$, where $T_o$ is a characteristic temperature representing the strength of carrier localization [23]. Figures 3(b) and 3(c) show plot of ln $\rho$ as a function of $T^{-1}$ and $T^{-1/4}$ and corresponding linear fits with thermal activation and VRH models, respectively. In a LT region, variation of each ln $\rho$ was linear but gradually deviated from the fitted lines with increasing temperature. The deviation started around 150 K for thermal activation model [Fig. 3(b)], while around 175 K for VRH model [Fig. 3(c)]. These results suggest that carrier conduction mechanism of the LT phase is governed by VRH associated with carrier localization at the W sites. Temperature dependence of Hall coefficient $R_H$ supported the carrier localization in LT phase as will be discussed later.

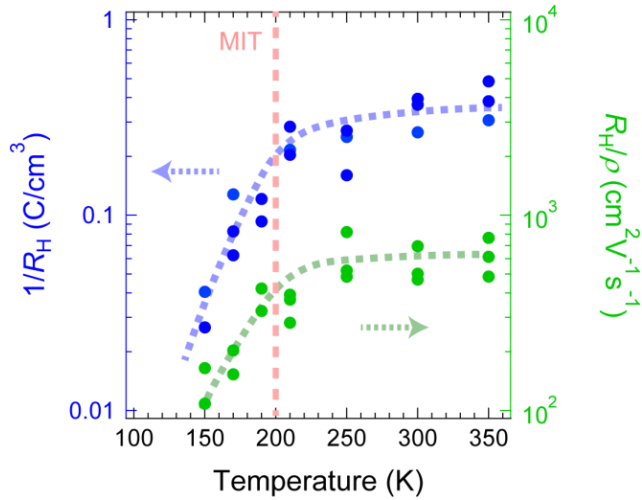

**Figure 4.** Temperature dependence of inverse Hall coefficient and Hall mobility. The blue and green broken lines and pink broken line indicate a guide to the eye and the transition temperature, respectively.

The magnetoresistance (MR) showed intriguing behavior across MIT and repeated sign reversal below 100 K. Figure 3(d) shows temperature dependence of MR of the CsW$_2$O$_6$ film taken at $H$ = 9 T, where MR($H$) is defined as $[\rho(H)-\rho(0)]/\rho(0)$. We subtracted odd function components from raw longitudinal $\rho$ data to obtain MR being an even function of $H$. The magnetic field dependence of MR taken at various temperatures are shown in Fig. 3(e) and Fig. 3(f). The negative MR was observed in a range of 120 K ≤ $T$ ≤ 300 K, where degree of MR modulated across the MIT. The larger signal of negative MR below 200 K suggests the enhancement of magnetic interaction in the LT insulating phase. The similar negative correlation between degree of MR and magnetization is generally observed for MIT materials [24]. On the other hand, finite negative MR was found even in the HT phase of our sample, implying that magnetic interaction still remained in the HT phase. In Ref. 9, the HT (LT) phase is considered as a paramagnetic metal (a nonmagnetic insulator). However, our results strongly suggest that both LT and HT phases near MIT are neither nonmagnetic nor paramagnetic.

At temperatures below 100 K, MR indicated anomalous sign reversal. As shown in Fig. 3 (f), the negative component gradually decreased with lowering temperature and disappeared around 60 K, where only positive MR with parabolic magnetic field dependence was found. In addition, the negative component revived at 40 K, implying another transition within the LT phase.

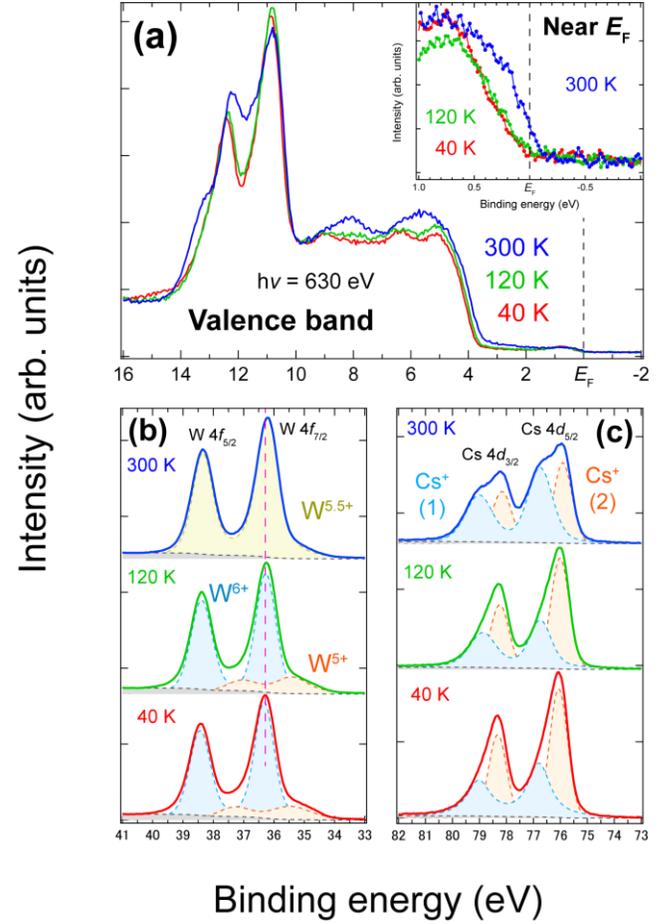

**Figure 5.** (a) Valence band, (b) W 4$f$ core-level, and (c) Cs 4$d$ core-level spectra of CsW$_2$O$_6$ films taken at 40, 120, and 300 K. The magnification near $E_F$ is shown in the inset of (a). The broken curves in (b) and (c) represent fitted results using Voigt functions.

Taking a half electron per W site and the indication of magnetic interactions into account, it is relevant to investigate characteristics of mobile charge carriers. The positive Hall coefficient $R_H$ with the linear slope as a function $H$ was revealed down to 100 K. Figure 4 shows temperature dependence of inverse $R_H$. The inverse $R_H$ suddenly decreased below 200 K, which was consistent with MIT. In contrast, the inverse $R_H$ in the HT phase was inde-pendent of temperature, which was also consistent with metallic nature. Hall mobility $R_H/\rho$ was ~600 cm$^2$V$^{-1}$s$^{-1}$ at 300 K, and it decreased below 200 K, reflecting VRH conduction, and reached ~100 cm$^2$V$^{-1}$s$^{-1}$ at 150 K. We suggest that the interaction between itinerant hole carriers and localized magnetic moment is responsible for the modulation of negative MR across MIT. On the other hand, fact that dominant charge carriers are holes is surprising, given broad bandwidth of the early transition-metal oxides.

To reveal the nature of the HT metallic and LT insulating phases, we performed synchrotron radiation PES across the MIT. Figure 5(a) shows the temperature dependence of valence band spectra for the CsW$_2$O$_6$ film. The density of states (DOS) around 14–10 eV was derived from the Cs $5p$ states, whose spectral shape changed with temperatures, which will be discussed later together with peak split of Cs $4d$ core levels. The broad DOS around 10–4 eV is mainly composed of the O $2p$ states, as it is common for other metal oxides [25]. The spectral shape of the O $2p$ states was slightly different between the HT (300 K) and LT (120 and 40 K) phases, reflecting the occurrence of the MIT associated with charge disproportionation. The small DOS near $E_F$ is associated with the W $5d$ states. In order to see the change of W $5d$ states between the HT and LT phases in detail, the magnification near $E_F$ is shown in the inset of Fig. 5(a). In the HT phase, small but finite DOS was found at $E_F$, which was a sign of metallic nature. In contrast, in the LT phase, a clear gap opened near $E_F$, which verified the insulating nature. These spectral features near $E_F$ indicate the MIT from the viewpoint of the electronic ground states and support the magneto-transport properties.

Figure 5(b) shows the temperature dependence of the W $4f$ core-level spectra. The W $4f$ core levels split into two peaks owing to spin-orbit interaction. The peaks centered at 36.5 (38.5) eV were assigned to the $J = 7/2$ (5/2) states. We deconvoluted the W $4f$ core-level spectra using Voigt functions. The fitted results are also indicated in Fig. 5(b). In the HT phase (300 K), the W $4f$ core-level spectrum was well fitted by only a single component, suggesting the delocalization of hole carriers residing in the W $5d$ state. Therefore, we tentatively assigned this profile to a W$^{5.5+}$ state and used as a reference for the following analysis. In contrast to the HT phase, W $4f$ core-level spectra for the LT phase (40 and 120 K) could not be fitted by another single component alone, but excess spectral weight could be derived from additional small component at the lower binding-energy side. The main component appeared at slightly higher binding energy compared with the reference W$^{5.5+}$ state. Although all the spectral features would have been smeared (to some extent) due to the surface oxidation layer and/or surface adsorbates, a chemical shift between those two components of ~ 1 eV coincides with that between W$^{6+}$ and W$^{5+}$ states [26–28]. Therefore, we assigned the main and small components to W$^{6+}$ and W$^{5+}$ states, respectively.

The spectral evolution of W $4f$ core levels suggests that the charge disproportionation from W$^{5.5+}$ to W$^{6+}$ and W$^{5+}$ occurs across the HT to LT phase transition. In fact, the W$^{5.5+}$ state was located in the middle of the W$^{6+}$ and W$^{5+}$ states as indicated by vertical broken line in Fig. 5(b). Such charge disproportionation and resultant spectral change are also reported for Spinel-type CuIr$_2$S$_4$ [29, 30]. CuIr$_2$S$_4$ with the pyrochlore lattice has $5d^{5.5}$ states (Ir$^{3.5+}$) with a low-spin configuration, which is electron-hole symmetry to $5d^{0.5}$ (W$^{5.5+}$) of CsW$_2$O$_6$ in degenerated $t_{2g}$ state. The single component (Ir$^{3.5+}$) of Ir $4f$ core-level spectrum split into two components (Ir$^{3+}$ and Ir$^{4+}$) below 226 K, where the MIT occurred.

Figure 5(c) shows the temperature dependence of the Cs $4d$ core-level spectra. Cs $4d$ core levels also split into two peaks owing to spin-orbit interactions. The peaks around 76 (78.5) eV were assigned to the $J = 5/2$ (3/2) states of Cs $4d$ core levels. Interestingly, their spectral shape drastically changed across the MIT despite the core levels of electronically inert alkali ion. Both Cs $4d$ core-level spectra for the LT-phase (40 and 120 K) were quite similar, suggesting that the spectral evolution did not merely follow decreasing temperature but associated with the MIT. To interpret anomalous spectral evolution, we also deconvoluted the Cs $4d$ core levels using Voigt functions. All the spectra were well fitted by two components assigned to Cs$^+$ (1) and Cs$^+$ (2) states. In the HT phase (300 K), the peak intensity of each component was comparable to the other. In contrast, those for the LT phase were significantly different to each other. We presume that Cs$^+$ (1) and Cs$^+$ (2) states are derived from the Cs$^+$ cations located at two different sites in the large cages surrounded by the octahedral WO$_6$ framework. The chemical shift of the Cs $4d$ core levels in principle arises from degree of core-hole screening of Cs$^+$ cations. The Cs$^+$ cations located at the farther (nearer) sites with respect to the cage W and O atoms would affect the smaller (larger) screening effects, resulting in the higher (lower) binding energy of the Cs $4d$ core level.

In the cubic $\beta$-pyrochlore lattice, $A^+$ cations in the large cages are known to change location, leading to rattling motion and multi-sites occupation [4]. In the HT phase of our sample, the Cs$^+$ cations likely occupy both Cs$^+$ (1) and Cs$^+$ (2) sites according to the rattling motion. On the other hand, Cs$^+$ cations may favor the nearer Cs$^+$ (2) sites in the LT phase owing to partial suppression of rattling motion associated with structural deformation from cubic to orthorhombic. Actually, it is revealed from the Rietveld analysis of powder X-ray and neutron diffractions that the Debye-Waller factor of the Cs atom significantly decreases in comparison to that of W and O atoms in the LT phase [9]. Furthermore, similar anomalous core-level spectra of rattling atoms are reported for cage materials [31-34], including a rattling superconducting material of Ba$_{24}$Ge$_{100}$, where the 12$d$-site Ba exists inside of the large cage of Ge. The 12$d$-site Ba $4d$ core levels split into two components (12$d$-1 and 12$d$-2) at 300 K reflecting the rattling motion. Moreover, their intensity ratio changes involving the lattice distortion at 20 K [30].

We discuss the aforementioned results to shed light on MIT and bad metallicity of the HT phase ($d\rho/dT < 0$), the latter of which is previously ascribed to poor crystallinity [9], and should now be considered as an intrinsic property of $CsW_2O_6$. The negative MR is clear indication of a correlated metal, where charge and/or spin fluctuations cause carrier scattering. In addition, rattling cations may interact with itinerate carries in the cages [4]. The charge disproportionation at the W site and suppression of rattling motion of $Cs^+$ cations in the LT phase manifest themselves across MIT, which involves the cubic-to-orthorhombic transition in bulk. Recent study for a rattling material of $Cu_{12}Sb_4S_{13}$ reveals that the MIT involving structural transition is attributed to $Cu^{2+}$ rattling displacements, and vice versa [35]. Therefore, manipulation of the rattling motion and charge disproportionate states of $CsW_2O_6$ is expected to influence MIT and even lead to the emergence of novel electronic phases. It has been recently pointed out that $CsW_2O_6$ can be superconducting upon suppressing the carrier localization [36]. Metallic and superconducting ground states are actually stabilized by applying high pressure or substitutional carrier doping in materials closely related to $CsW_2O_6$ ($Cu_{12}Sb_4S_{13}$ [37] and $CuIr_2S_4$ [38]). Furthermore, because of thin film form, epitaxial strain and ionic liquid-gated carrier accumulation, which were recently developed as powerful tools for controlling physical properties, were also usable. Using these advanced techniques, the LT insulating phase could be suppressed and exotic electronic phases would be induced in $CsW_2O_6$.

## ■ CONCLUSION

In summary, we have successfully grown single-crystalline $\beta$–pyrochlore-type $CsW_2O_6$ epitaxial films on YSZ (111) substrates using the PLD technique. The temperature dependence of resistivity clearly indicated the phase transition at 200 K. The HT phase exhibited rather insulating behavior in the temperature dependence of resistivity, indicating intrinsic bad metallicity. The magnetotransport measurements revealed negative magnetoresistances and positive Hall coefficient in both LT and HT phases. It was revealed from PES measurements that the valence band spectra indicated DOS at $E_F$ in the HT metallic phase, which vanished in the LT insulating phase. The W $4f$ core-level spectrum changed its shape across the MIT, where two components were clearly discriminated in the insulating phase, suggesting the charge disproportionation of $W^{5+}$ and $W^{6+}$. The Cs $4d$ core-level spectrum also drastically changed its spectral shape involving the MIT, reflecting the rattling motion of $Cs^+$ cations. The magnetotransport and photoemission measurements strongly suggest that $CsW_2O_6$ is the first material, which shows the charge disproportionate MIT associated with rattling motion.

## ■ AUTHOR INFORMATION

**Corresponding Author**

*E-mail: k-yoshi@apc.titech.ac.jp

**Notes**

The authors declare no competing financial interest.

## ■ ACKNOWLEDGMENT

This work was partly supported by MEXT Elements Strategy Initiative to Form Core Research Center and a Grant-in-Aid for Scientific Research (Nos. 16H05983, 18H03925, and 18J10171). The work at KEK-PF was done under the approval of the Program Advisory Committee (Proposals Nos. 2017G596 and No. 2015S2-005) at the Institute of Materials Structure Science, KEK. T. S. acknowledges the financial support from JSPS Research Fellowship for Young Scientists.

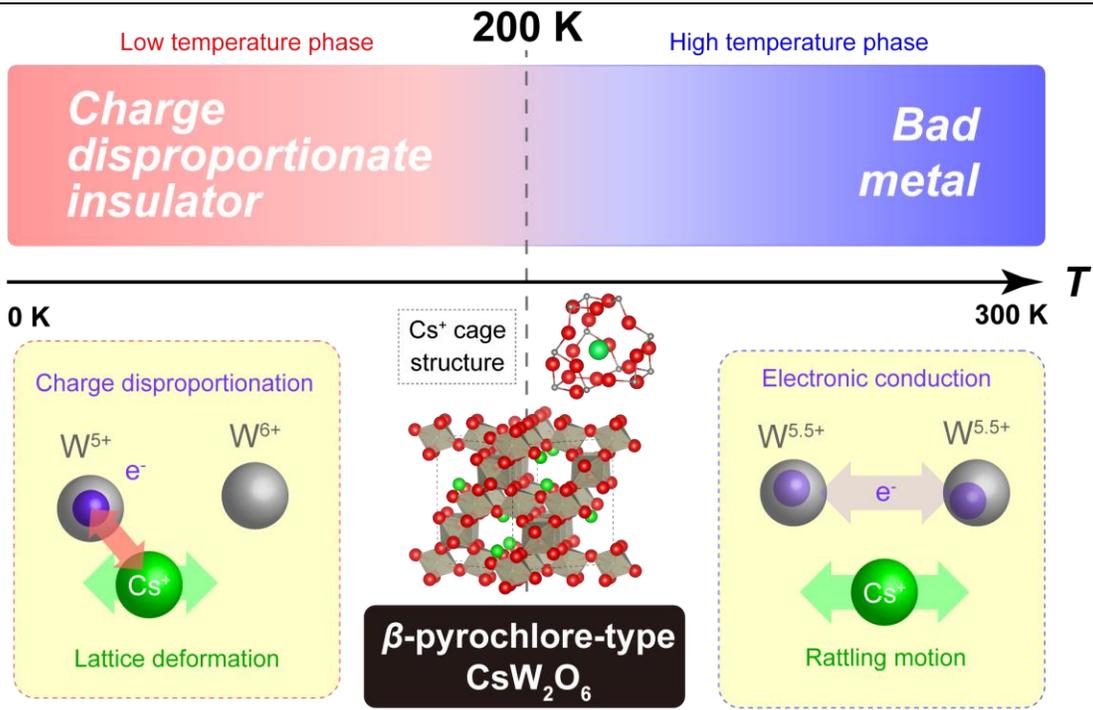